\documentclass[showpacs,preprintnumbers,prd,nofootinbib,floats,amssymb,floatfix]{revtex4-2}
\usepackage{amsmath}
\usepackage{graphicx}
\usepackage{amsxtra}
\usepackage{hyperref}
\usepackage{amssymb}
\usepackage{amstext}
\usepackage{amsmath}
\usepackage{cleveref}
\usepackage{blkarray}
\begin{document}

\title{Canonical analysis for Chern-Simons modification of  general relativity}
\author{Alberto Escalante}  \email{aescalan@ifuap.buap.mx}
\author{J. Aldair Pantoja-Gonz\'alez }  \email{jpantoja@ifuap.buap.mx}
 \affiliation{Instituto de F\'isica, Benem\'erita Universidad Aut\'onoma de Puebla. \\ Apartado Postal J-48 72570, Puebla Pue., M\'exico, }
\begin{abstract}
By using the Gitman-Lyakhovich-Tyutin  canonical analysis
for higher-order theories  a four-dimensional Chern-Simons modification of general relativity is analyzed. The counting of physical
degrees of freedom, the symmetries, and the fundamental Dirac brackets are reported. Additionally, we report the complete structure of the constraints and its  Dirac algebra is developed.
\end{abstract}
 \date{\today}
\preprint{}
\maketitle
\section{Introduction}
It is well known that the addition of topological terms to physical theories presents contributions that could change the  physical description of the base theory. In fact, there are  those  that do not affect  the equations of motion, being total derivatives or identities; these theories do not modify the degrees of freedom, but modify the symplectic structure,  and those that contribute to the equations of motion where the topological terms are coupled  with coordinate-dependent parameters.  For instance, the addition of the topological Chern-Simons [CS] term to three-dimensional Maxwell theory  turns   the gauge fields  into massive ones, modifying the dynamics due to the presence of an extra degree of freedom \cite{1, 2, 4}. In the four-dimensional case, the coupling of the Pontryagin invariant to Maxwell's theory  modifies the photon physics \cite{5}. In fact, the dynamics are modified, now the photon retains two degrees of freedom, but in vacuum Lorentz invariance and parity are lost. In the case of theories describing gravity, there are also interesting scenarios in which   topological theories or topological terms play  an interesting role in the physical degrees of freedom of the gravitational field. In the tree dimensional case, the coupling of the CS invariant to Einstein Hilbert  [EH] action provides mass to the  graviton  and modifies the symplectic structure of theory \cite{3}. Moreover, in the four-dimensional stage,  the Macdowell–Mansouri formulation is worth mentioning. In fact, that  formulation   breaks down the symmetry of a BF theory from SO(5) to SO(4)  obtaining the Palatini action plus the sum of topological invariants \cite{6, 7, 8};  thus, the breaking symmetry of a topological theory leads to new physical degrees of freedom. On the other hand, in the canonical gravity context, there is also the well-known Holst action \cite{9}. This action is given by the Palatini action plus a topological term, it depends on the so-called Barbero–Immirizi [BI] parameter and provides a set of actions classically equivalent to Einstein’s theory. In fact, both Palatini and Holst actions share the same equations of motion, however, the topological term through the BI parameter  contributes at the classical level in the symplectic structure of the theory. Furthermore, by coupling  the Holst action with fermionic mater,  the BI parameter determines the coupling constant interaction between fermions \cite{10}. From the
quantum point of view, this  parameter gives a contribution
in the quantum spectra of the area and volume operators in
the context of loop quantum gravity \cite{11, 12, 13}. Moreover, the Holst term facilitates the
canonical description of gravity, and depending on the values of the BI parameter one can reproduce the different scenarios found in canonical gravity. \\
On the other hand, there is a model found in the literature proposed by  R. Jackiw and Y. Pi [JY] where the topological Pontryagin term is coupled through an auxiliary field to   Einstein-Hilbert [EH] action \cite{Jackiw}. Under a particular configuration of the external field,  this theory presents some interesting features: the Schwarzschild metric is also a solution of the modified theory, thus, there are no changes in the basic predictions of general relativity. Moreover, the theory describes the propagation of gravitational waves with velocity $c$, but these  carry different intensities and violate spatial reflection symmetry. However, the analysis reported  in \cite{Jackiw} was performed  only at Lagrangian level and as we know any modification could change the canonical structure of the base theory or the number  of physical degrees of freedom.    In this respect,  it is mandatory to perform a canonical analysis in order to determine  if there are any  changes in the canonical structure of the constraints in the modified theory, keeping in mind that the constraints are the best guideline to perform a canonical quantization. Furthermore,  there are some examples where the Hamiltonian  and Lagrangian degrees of freedom counting disagree,  for instance  the higher-order theory studied in  \cite{sk}, thus, although a Lagrangian study  of  the $JY$ theory has been reported, the Hamiltonian analysis  is important. \\
With all this being said,  the objective  of this paper is to present the canonical analysis of the theory reported in \cite{Jackiw}. Our study will be developed in the perturbative context  around the Minkowski spacetime, this particular choice  will show that the JY theory is a higher-order singular theory in the temporal derivatives. Moreover, it is worth commenting that the canonical analysis of gauge higher-order theories is a difficult task to develop and modern tools for performing the analysis are needed. In this respect, there are two approaches for performing the analysis of higher-order gauge theories the so-called Ostrogradski-Dirac [OD] and Gitman–Lyakhovich–Tyutin [GLT] methods \cite{15, 16, 16a}. The former is based on the extension of the phase space, where the choice of the fields
and their temporal derivatives become  the canonical variables, thus, a generalization of the canonical momenta for the higher-temporal derivative of the fields is introduced. However,
it is claimed that in some cases the usual OD framework does not allow an easy identification of  the complete structure of the constraints, then the constraints are fixed by hand in order to achieve  consistency \cite{17}. On the other hand, the GLT framework is a generalization of Ostrogradski’s method,  based on the introduction of extra fields reducing a
problem with higher temporal derivatives to one with first-order time
derivatives.  Then by using either the definition of the momenta for all the fields or
the introduction of the Dirac brackets, the second class constraints, and non-physical degrees of freedom (extra fields) can be removed \cite{18, 19}. In this manner, we will utilize  the GLT formalism to perform our analysis due to it is sturdiness at handling the constraints  and their consistency. In fact, we will report that there exists a contribution to the canonical structure of the EH theory due to the CS modification. We will report the complete structure of the constraints and  we will perform the counting of degrees of freedom, then we will remove the non-physical degrees of freedom by introducing the fundamental Dirac-brackets  and the Dirac algebra between all constraints will be reported. \\
The paper is organized as follows. In Section II we present a brief description of the JY theory, then the GLT analysis is performed. We will report the complete canonical structure of the theory, namely,  the complete structure of the constraints, the counting of physical degrees of freedom is developed and the fundamental Dirac brackets are constructed. In Section III we present the conclusions.
\section{The Gitman-Lyakhovich-Tyutin analysis.}
We start with  the following action composed by  the EH action plus  the Pontryagin invariant \cite{Jackiw}, this is 
\begin{equation}
\label{EH+CS}
S[g_{\mu\nu}]=\int\left(R\sqrt{-g}+\frac{1}{4 }\theta {*}R^{\sigma}{}_{\tau}{}^{\mu\nu}R^{\tau}{}_{\sigma\mu\nu}\right)d^{4}x,
\end{equation}
where $g_{\mu\nu}$ is the metric tensor, $\theta$ is a coupling field which in general is not constant and ${*}R^{\sigma}{}_{\tau}{}^{\mu\nu}= \frac{1}{2} \varepsilon^{\mu\nu\alpha\beta}R^{\sigma}{}_{\tau\alpha\beta}$. Along this paper we will use Greek letters  for labeling the spacetime  indices  $\mu, \nu,..,\alpha=0,...,3$ and latin ones  $i, j..,k=1, 2, 3$ for labeling space indices. It is worth mentioning, that the action (\ref{EH+CS}) is called a non-dynamical theory, however,  there exist other generalizations where one can include matter fields or scalar fields, then the theory is called dynamical one.  These scenarios of the dynamical theory  are  relevant in particle physics and string theory due to the effective theory  presenting   anomaly cancelation, for an extended  review on CS modified gravity can be consulted in the reference \cite{Alexander}.  Furthermore, we can rewrite the action by noticing that
\begin{eqnarray}
\frac{1}{2}{*}R^{\sigma}{}_{\tau}{}^{\mu\nu}R^{\tau}{}_{\sigma\mu\nu}&=&
 2\varepsilon^{\mu\alpha\beta\nu}\partial_{\mu}\left(\frac{1}{2}\Gamma^{\sigma}_{\alpha\tau}\partial_{\beta}\Gamma^{\tau}_{\nu\sigma} + \frac{1}{3}\Gamma^{\sigma}_{\alpha\tau}\Gamma^{\tau}_{\beta\eta}\Gamma^{\eta}_{\nu\sigma} \right)
\\
\nonumber
&=& \partial_{\mu}K^{\mu},
\end{eqnarray}
where $K^{\mu}=2\epsilon^{\mu\alpha\beta\gamma}\left[\frac{1}{2}\Gamma_{\alpha\tau}^{\sigma}\partial_{\beta}\Gamma_{\gamma\sigma}^{\tau}+\frac{1}{3}\Gamma_{\alpha\tau}^{\sigma}\Gamma_{\beta\eta}^{\tau}\Gamma_{\gamma\sigma}^{\eta}\right]$. Then, up to a boundary term  the action takes the form
\begin{equation}
S[g_{\mu\nu}]=\int\left(R\sqrt{-g}-\frac{1}{2}\mathfrak{v}_{\mu}K^{\mu}\right)d^{4}x, 
\label{eq3}
\end{equation}
where  $\mathfrak{v}_{\mu}\equiv\partial_{\mu}\theta$. The variation of the action (\ref{eq3}) yields the following equations of motion
\begin{equation}
\label{G+C}
\mathcal{G}^{\mu\nu}+\mathcal{C}^{\mu\nu}=0,
\end{equation}
here $\mathcal{G}^{\mu\nu}$ is the Einstein tensor and $\mathcal{C}^{\mu\nu}\equiv-\frac{1}{2\sqrt{-g}}[\mathfrak{v}_{\sigma}(\epsilon^{\sigma\mu\alpha\beta}D_{\alpha}R^{\nu}{}_{\beta}+\epsilon^{\sigma\nu\alpha\beta}D_{\alpha}R^{\mu}{}_{\beta})+\mathfrak{v}_{\sigma\tau}({}^{*}R^{\tau\mu\sigma\nu}+{}^{*}R^{\tau\nu\sigma\mu})]$ is a  four dimensional Cotton-type tensor and $\mathfrak{v}_{\mu\nu}=\partial_{\mu}\partial_{\nu}\theta$. We can fix  a particular form of the coupling field, the so-called canonical CS coupling, namely 
\begin{equation}
\label{CanonicalCoupling}
\theta=\frac{t}{\Omega} \quad \longleftrightarrow \quad \mathfrak{v}_{\mu}=(1/\Omega,0,0,0),
\end{equation}
 thus, the Schwarzchild metric is also a solution of  (\ref{G+C}) \cite{Jackiw, Alexander}  and the classical test of GR are

  considered  in the modification. In this manner, we are interested in performing  a canonical analysis of the action (\ref{eq3}) in order to find any contribution of the CS term to the well-known canonical structure of GR.  For our aims, we shall find the linearized  action around the Minkowski spacetime by using the usual perturbation 
\begin{equation}
\label{WeakField}
g^{\mu\nu}=\eta^{\mu\nu}- h^{\mu\nu}(x),
\end{equation}
hence, by substituting  \eqref{CanonicalCoupling},  \eqref{WeakField} into \eqref{EH+CS} and performing integration by parts we get 
\begin{eqnarray}
\nonumber
S[h_{\mu\nu}] &=& \int\left(\sqrt{-g}R-\frac{1}{2}\mathfrak{v}_{\mu}K^{\mu}\right)d^{4}x
\\
\nonumber
&=& \int \left[\left(\frac{1}{4}\partial_{\lambda} h_{\mu\nu}\partial^{\lambda}h^{\mu\nu}-\frac{1}{4}\partial_{\lambda} h^{\mu}{}_{\mu}\partial^{\lambda}h^{\nu}{}_{\nu}+\frac{1}{2}\partial_{\lambda}h^{\lambda}{}_{\mu}\partial^{\mu}h^{\nu}{}_{\nu}-\frac{1}{2}\partial_{\lambda}h^{\lambda}{}_{\mu}\partial_{\nu}h^{\nu\mu}\right)\right.
\\
\label{EH+CS 2}
&& - \frac{1}{4\Omega}\epsilon^{0\lambda\mu\nu}\left.\bigg(\partial_{\sigma}h_{\lambda}{}^{\rho}\partial_{\rho}\partial_{\mu}h_{\nu}{}^{\sigma}-\partial_{\sigma}h_{\lambda}{}^{\rho}\partial^{\sigma}\partial_{\mu}h_{\rho\nu}\bigg)\right]d^{4}x,
\end{eqnarray}
in the second line we also neglect higher-order terms in $h_{\mu\nu}$, the  variation of \eqref{EH+CS 2} yields 
\begin{equation}
\label{LinearEquations}
\mathcal{G}_{\mu\nu}^{lin}+\mathcal{C}_{\mu\nu}^{lin}=0,
\end{equation}
where
\begin{eqnarray}
\nonumber
\mathcal{G}_{\mu\nu}^{lin}&=&\frac{1}{2}[\square h_{\mu\nu}+\partial_{\mu}\partial_{\nu}h^{\lambda}{}_{\lambda}-\partial_{\mu}\partial_{\lambda}h^{\lambda}{}_{\nu}-\partial_{\nu}\partial_{\lambda}h^{\lambda}{}_{\mu}-\eta_{\mu\nu}(\square h^{\lambda}{}_{\lambda}-\partial_{\lambda}\partial_{\gamma}h^{\lambda\gamma}]],
\\
\mathcal{C}_{\mu\nu}^{lin}&=&-\frac{1}{4\Omega}[\epsilon_{0\mu\lambda\gamma}\partial^{\lambda}(\square h^{\gamma}{}_{\nu}-\partial_{\nu}\partial_{\alpha}h^{\alpha\gamma})+\epsilon_{0\nu\lambda\gamma}\partial^{\lambda}(\square h^{\gamma}{}_{\mu}-\partial_{\mu}\partial_{\alpha}h^{\alpha\gamma})],
\label{eqlin}
\end{eqnarray}
here $\Box$ is the D'Alambertian operator. Furthermore,  we can observe that the action (\ref{EH+CS 2}) can be written as
\begin{equation}
\label{I}
S[h_{\mu\nu}]=-\frac{1}{2}\int h^{\mu\nu}\left(\mathcal{G}_{\mu\nu}^{lin}+\mathcal{C}_{\mu\nu}^{lin}\right),
\end{equation}
hence, this new fashion of the action   (\ref{EH+CS 2}) will be analyzed bellow. 
 As far as we know, the canonical analysis of the action (\ref{I})  has not been reported in the literature.\\
We will remove the  overall  factor of $-\frac{1}{2}$ from  (\ref{I}), which of  course, does not change the dynamics. Moreover, we  perform the  $3+1$ decomposition of action \eqref{I},  thus, the  action takes the following form
\begin{eqnarray}
\nonumber
S&=&\int\left[\frac{1}{2}\dot{h}_{ij}\dot{h}^{ij}-\partial_{j}h_{0i}\partial^{j}h^{0i}-\frac{1}{2}\partial_{k}h_{ij}\partial^{k}h^{ij}-\frac{1}{2}\dot{h}^{i}{}_{i}\dot{h}^{j}{}_{j}+\partial^{j}h^{0}{}_{0}\partial_{j}h^{i}{}_{i}+\frac{1}{2}\partial_{k}h^{i}{}_{i}\partial^{k}h^{j}{}_{j}\right.
\\
\nonumber
& & -2\partial^{i}h^{0}{}_{i}\dot{h}^{j}{}_{j}-\partial_{i}h^{0}{}_{0}\partial_{j}h^{ij}-\partial_{i}h^{ij}\partial_{j}h^{k}{}_{k}+2\partial_{j}h^{0}{}_{i}\dot{h}^{ij}+\partial_{i}h^{i}{}_{0}\partial_{j}h^{0j}+\partial_{k}h^{k}{}_{i}\partial_{j}h^{ij}
\\
\label{Action1}
& & \left. +\frac{1}{\mu}\epsilon^{0ijk}(-\ddot{h}^{l}{}_{i}\partial_{j}h_{lk}+2\dot{h}^{l}{}_{i}\partial_{j}\partial_{l}h^{0}{}_{k}+\partial_{l}h^{m}{}_{i}\partial_{m}\partial_{j}h^{l}{}_{k}+\nabla^{2}h^{0}{}_{i}\partial_{j}h_{0 k}+\nabla^{2}h^{m}{}_{i}\partial_{j}h_{m k})\right]d^{4}x,
\end{eqnarray}
where we have defined $\mu\equiv2\Omega$. We can observe that the action is a higher-order theory and we will use the GLT method for performing the canonical analysis.   By following  the GLT framework (see the appendix A),  we introduce the following   set of variables \cite{18, 19} 
\begin{eqnarray}
\label{ExtraFields1}
G_{\mu\nu}&\equiv&\dot{h}_{\mu\nu},
\\
\label{ExtraFields2}
v_{\mu\nu}&\equiv&\ddot{h}_{\mu\nu},
\end{eqnarray}
hence, by introducing these  new variables through Lagrange multipliers, the action  \eqref{Action1} will take the form  
\begin{equation}
\label{Lagrangian}
\mathcal{S}'=\int \mathcal{L'}dx^4 =\int \mathcal{L} dx^4+\int[\lambda_{1}^{\mu\nu}(\dot{h}_{\mu\nu}-G_{\mu\nu})+\lambda_{2}^{\mu\nu}(\dot{G}_{\mu\nu}-v_{\mu\nu})]d^{4}x,
\end{equation}
where 
\begin{eqnarray}
\nonumber
\mathcal{L}&=&\int\left[\frac{1}{2}G_{ij}G^{ij}-\partial_{j}h_{0i}\partial^{j}h^{0i}-\frac{1}{2}\partial_{k}h_{ij}\partial^{k}h^{ij}-\frac{1}{2}G^{i}{}_{i}G^{j}{}_{j}+\partial^{j}h^{0}{}_{0}\partial_{j}h^{i}{}_{i}+\frac{1}{2}\partial_{k}h^{i}{}_{i}\partial^{k}h^{j}{}_{j}\right.
\\
\nonumber
& & -2\partial^{i}h^{0}{}_{i}G^{j}{}_{j}-\partial_{i}h^{0}{}_{0}\partial_{j}h^{ij}-\partial_{i}h^{ij}\partial_{j}h^{k}{}_{k}+2\partial_{j}h^{0}{}_{i}G^{ij}+\partial_{i}h^{i}{}_{0}\partial_{j}h^{0j}+\partial_{k}h^{k}{}_{i}\partial_{j}h^{ij}
\\
\nonumber
& & \left. +\frac{1}{\mu}\epsilon^{0ijk}(-v^{l}{}_{i}\partial_{j}h_{lk}+2G^{l}{}_{i}\partial_{j}\partial_{l}h^{0}{}_{k}+\partial_{l}h^{m}{}_{i}\partial_{m}\partial_{j}h^{l}{}_{k}+\nabla^{2}h^{0}{}_{i}\partial_{j}h_{0 k}+\nabla^{2}h^{m}{}_{i}\partial_{j}h_{m k})\right]d^{3}x,
\end{eqnarray}
We now introduce the momenta $(\pi^{\mu\nu},p^{\mu\nu},\hat{v}^{\mu\nu},\Lambda_{\mu\nu}^{1},\Lambda_{\mu\nu}^{2})$ canonically conjugate to $(h_{\mu\nu},G_{\mu\nu},v_{\mu\nu},\lambda_{1}^{\mu\nu},\lambda_{2}^{\mu\nu})$, these are 
\begin{eqnarray}
\label{Momenta1}
\pi^{\mu\nu}&\equiv&\frac{\partial\mathcal{L}'}{\partial\dot{h}_{\mu\nu}}=\lambda_{1}^{\mu\nu},
\\
\label{Momenta2}
p^{\mu\nu}&\equiv&\frac{\partial\mathcal{L}'}{\partial\dot{G}_{\mu\nu}}=\lambda_{2}^{\mu\nu},
\\
\label{Momenta3}
\hat{v}^{\mu\nu}&\equiv&\frac{\partial\mathcal{L}'}{\partial\dot{v}_{\mu\nu}}=0,
\\
\label{Momenta4}
\Lambda_{\mu\nu}^{1}&\equiv&\frac{\partial\mathcal{L}'}{\partial\dot{\lambda}_{1}^{\mu\nu}}=0,
\\
\label{Momenta5}
\Lambda_{\mu\nu}^{2}&\equiv&\frac{\partial\mathcal{L}'}{\partial\dot{\lambda}_{2}^{\mu\nu}}=0.
\end{eqnarray}
We observe that the equations \eqref{Momenta1}--\eqref{Momenta2} allows us to identify the Lagrange multipliers  $(\lambda_{1}^{\mu\nu},\lambda_{2}^{\mu\nu})$ as canonical variables through the momenta $(\pi^{\mu\nu},p^{\mu\nu} )$ respectively. Additionally,  the primary constraints will be given by \cite{18}
\begin{equation}
\varphi^{\mu\nu}\equiv p^{\mu\nu}-\frac{\partial\mathcal{L}'}{v_{\mu\nu}}, 
\end{equation}
in this manner, we identify the following primary constraints 
\begin{eqnarray}
\nonumber
\varphi^{00}&\equiv&p^{00}\approx0,
\\
\nonumber
\varphi^{0i}&\equiv&p^{0i}\approx0,
\\
\label{PrimaryConstraints}
\varphi^{ij}&\equiv&p^{ij}+\frac{1}{2\mu}(\epsilon^{ikl}\eta^{jm}+\epsilon^{jkl}\eta^{im})\partial_{k}h_{lm}\approx0, 
\end{eqnarray}
and the fundamental Poisson brackets of the theory will be  expressed as 
\begin{eqnarray}
\label{FundamentalBrackets1}
\lbrace h_{\mu\nu},\pi^{\alpha\beta} \rbrace &=& \frac{1}{2}(\delta_{\mu}^{\alpha}\delta_{\nu}^{\beta}+\delta_{\mu}^{\beta}\delta_{\nu}^{\alpha})\delta^{3}(x-y),
\\
\label{FundamentalBrackets2}
\lbrace G_{\mu\nu},p^{\alpha\beta} \rbrace &=& \frac{1}{2}(\delta_{\mu}^{\alpha}\delta_{\nu}^{\beta}+\delta_{\mu}^{\beta}\delta_{\nu}^{\alpha})\delta^{3}(x-y).
\end{eqnarray}
With all  momenta identified  we define  the canonical Hamiltonian given by 
\begin{eqnarray}
\nonumber
\mathcal{H}_{can}&=&\int[\pi^{\mu\nu}G_{\mu\nu}+p^{\mu\nu}v_{\mu\nu}]d^{3}x-\int[\frac{1}{2}G_{ij}G^{ij}-\partial_{j}h_{0i}\partial^{j}h^{0i}-\frac{1}{2}\partial_{k}h_{ij}\partial^{k}h^{ij}-\frac{1}{2}G^{i}{}_{i}G^{j}{}_{j}-\partial^{j}h_{00}\partial_{j}h^{i}{}_{i}
\\
\nonumber
&+&\frac{1}{2}\partial_{k}h^{i}{}_{i}\partial^{k}h^{j}{}_{j}+2\partial^{i}h_{0i}G^{j}{}_{j}+\partial_{i}h_{00}\partial_{j}h^{ij}-\partial_{i}h^{ij}\partial_{j}h^{k}{}_{k}-2\partial_{j}h_{0i}G^{ij}-\partial^{i}h_{0i}\partial^{j}h_{0j}+\partial_{k}h^{k}{}_{i}\partial_{j}h^{ij}
\\
\label{CanonicalHamiltonian}
&+&\frac{1}{\mu}\epsilon^{ijk}(-v^{l}{}_{i}\partial_{j}h_{lk}-2G^{l}{}_{i}\partial_{j}\partial_{l}h_{0k}+\partial_{l}h^{m}{}_{i}\partial_{m}\partial_{j}h^{l}{}_{k}-\nabla^{2}h_{0i}\partial_{j}h_{0 k}+\nabla^{2}h^{m}{}_{i}\partial_{j}h_{m k})]d^{3}x, 
\end{eqnarray}
thus,  the primary Hamiltonian  will be expressed  in the following form
\begin{equation}
\mathcal{H}_{1}=\mathcal{H}_{can}+\int \Delta_{\mu\nu}\varphi^{\mu\nu}d^{3}x,
\end{equation}
where $\Delta_{\mu\nu}$ are Lagrange multipliers enforcing the  primary constraints. It is worth mentioning, that the canonical Hamiltonian presents linear terms in the momenta, thus, this fact could be associated to Ostrogradski's instabilities. However, we will see at the end of the paper that this apparently instability can be healed by  introducing the Dirac brackets, we have added the appendix B for clarifying this point. 
 In order to identify  further  constraints,   we will calculate the consistency conditions of the primary constraints. From the consistency conditions  we obtain the following secondary constraints
\begin{eqnarray}
\label{SecondaryConstraints}
\nonumber
\dot{\varphi}^{00} &=& \lbrace \varphi^{00},\mathcal{H}_{1} \rbrace  \approx 0,
\\
&& \Rightarrow \Phi^{00}\equiv\pi^{00}\approx0,
\\
\nonumber
\label{27a}
\dot{\varphi}^{0i} &=& \lbrace \varphi^{0i},\mathcal{H}_{1} \rbrace  \approx 0,
\\
&&  \Rightarrow \Phi^{0i}\equiv\pi^{0i}\approx0,
\\
\nonumber
\label{28a}
\dot{\varphi}^{ij} &=& \lbrace \varphi^{ij},\mathcal{H}_{1} \rbrace  \approx 0, 
\\ \nonumber
&&\Rightarrow \Phi^{ij}\equiv\pi^{ij}+\frac{1}{\mu}(\epsilon^{ikl}\partial^{j}+\epsilon^{jkl}\partial^{i})\partial_{k}h_{0l}-\frac{1}{2\mu}(\epsilon^{ikl}\eta^{jm}+\epsilon^{jkl}\eta^{im})\partial_{k}G_{lm}
\\
& & -\:G^{ij}+(G^{k}{}_{k}-2\partial^{k}h_{0k})\eta^{ij}+(\partial^{i}h_{0}{}^{j}+\partial^{j}h_{0}{}^{i})\approx0,
\end{eqnarray}
Thus, from consistency  of  the secondary constraints (\ref{SecondaryConstraints}), (\ref{27a}) and (\ref{28a})  we obtain 
\begin{eqnarray}
\nonumber
\dot{\Phi}^{00} &=& \lbrace \Phi^{00},\mathcal{H}_{1} \rbrace  \approx 0,
\\
&& \Rightarrow \nabla^{2}h^{i}{}_{i} - \partial_{i}\partial_{j}h^{ij} \approx 0,
\\
\nonumber
\\
\nonumber
\dot{\Phi}^{0i} &=& \lbrace \Phi^{0i},\mathcal{H}_{1} \rbrace \approx 0,
\\
&& \Rightarrow \frac{1}{\mu}\epsilon^{ijk}(\partial_{j}\partial^{l}G_{kl}-\nabla^{2}\partial_{j}h_{0k})-\nabla^{2}h_{0}{}^{i}-\partial^{i}G^{j}{}_{j}+\partial_{j}G^{ij}+\partial^{i}\partial^{j}h_{0j} \approx 0,
\\
\nonumber
\\
\nonumber
\dot{\Phi}^{ij} &=& \lbrace \Phi^{ij},\mathcal{H}_{1} \rbrace \approx 0,
\\
\nonumber
&& \Rightarrow
\frac{1}{\mu}[(\epsilon^{ikl}\partial^{j}+\epsilon^{jkl}\partial^{i})\partial_{k}G_{0l}+(\epsilon^{ikl}\eta^{jm}+\epsilon^{jkl}\eta^{im})\nabla^{2}\partial_{k}h_{lm}-(\epsilon^{ikl}\partial^{j}+\epsilon^{jkl}\partial^{i})\partial^{m}\partial_{k}h_{lm}
\\
\nonumber
& & - \: (\epsilon^{ikl}\eta^{jm}+\epsilon^{jkl}\eta^{im})\partial_{k}v_{lm}]+\frac{1}{2\mu}(\epsilon^{ikm}\eta^{jl}+\epsilon^{jkm}\eta^{il}+\epsilon^{ilm}\eta^{jk}+\epsilon^{jlm}\eta^{ik})\partial_{m}\Delta_{kl}
\\
\nonumber
& & + \: \nabla^{2}h^{ij}-\partial^{i}\partial^{j}h_{00}+\partial^{i}\partial^{j}h^{k}{}_{k}-(\partial^{i}\partial_{k}h^{jk}+\partial^{j}\partial_{k}h^{ik})+(\nabla^{2}h_{00}-\nabla^{2}h^{k}{}_{k}+\partial_{k}\partial_{l}h^{kl})\eta^{ij}
\\
& & + \: (\partial^{i}G_{0}{}^{j}+\partial^{j}G_{0}{}^{i})-2\partial^{k}G_{0k}\eta^{ij} -  v^{ij} + v^{k}{}_{k}\eta^{ij}-\Big[\frac{1}{2}(\eta^{ik}\eta^{jl}+\eta^{jk}\eta^{il})-\eta^{ij}\eta^{kl}\Big]\Delta_{kl} \approx 0,
\end{eqnarray}
where we  identify the following tertiary constraints
\begin{eqnarray}
\label{32a}
\Psi^{00}&\equiv&\nabla^{2}h^{i}{}_{i}-\partial_{i}\partial_{j}h^{ij},
\\
\label{33a}
\Psi^{0i}&\equiv&\frac{1}{\mu}\epsilon^{ijk}(\partial_{j}\partial^{l}G_{kl}-\nabla^{2}\partial_{j}h_{0k})-\nabla^{2}h_{0}{}^{i}+\partial^{i}\partial^{j}h_{0j}-\partial^{i}G^{j}{}_{j}+\partial_{j}G^{ij},
\end{eqnarray}
and the following relations for the multipliers $\Delta_{ij}$
\begin{eqnarray}
\nonumber
&& \frac{1}{\mu}[(\epsilon^{ikl}\partial^{j}+\epsilon^{jkl}\partial^{i})\partial_{k}G_{0l}+(\epsilon^{ikl}\eta^{jm}+\epsilon^{jkl}\eta^{im})\nabla^{2}\partial_{k}h_{lm}-(\epsilon^{ikl}\partial^{j}+\epsilon^{jkl}\partial^{i})\partial^{m}\partial_{k}h_{lm}
\\
\nonumber
&& - \: (\epsilon^{ikl}\eta^{jm}+\epsilon^{jkl}\eta^{im})\partial_{k}v_{lm}]+\frac{1}{2\mu}(\epsilon^{ikm}\eta^{jl}+\epsilon^{jkm}\eta^{il}+\epsilon^{ilm}\eta^{jk}+\epsilon^{jlm}\eta^{ik})\partial_{m}\Delta_{kl}
\\
\nonumber
&& + \: \nabla^{2}h^{ij}-\partial^{i}\partial^{j}h_{00}+\partial^{i}\partial^{j}h^{k}{}_{k}-(\partial^{i}\partial_{k}h^{jk}+\partial^{j}\partial_{k}h^{ik})+(\nabla^{2}h_{00}-\nabla^{2}h^{k}{}_{k}+\partial_{k}\partial_{l}h^{kl})\eta^{ij}
\\
\label{Multipliers1}
& & + \: (\partial^{i}G_{0}{}^{j}+\partial^{j}G_{0}{}^{i})-2\partial^{k}G_{0k}\eta^{ij} -  v^{ij} + v^{k}{}_{k}\eta^{ij}-\left[\frac{1}{2}(\eta^{ik}\eta^{jl}+\eta^{jk}\eta^{il})-\eta^{ij}\eta^{kl}\right]\Delta_{kl}=0.
\end{eqnarray}
Now, we calculate  consistency on the  tertiary constraints (\ref{32a}) and (\ref{33a}), we find 
\begin{eqnarray}
\nonumber
\dot{\Psi}^{00} &=& \lbrace \Psi^{00},\mathcal{H}_{1} \rbrace \approx 0,
\\
&& \Rightarrow
\nabla^{2}G^{i}{}_{i} - \partial^{i}\partial^{j}G_{ij}= \partial_i \Psi^{0i}=0,
\\
\nonumber
\end{eqnarray}
\begin{eqnarray}
\dot{\Psi}^{0i}&=& \lbrace \Psi^{0i},\mathcal{H}_{1} \rbrace \approx 0,
\nonumber \\
&& \Rightarrow
\frac{1}{\mu}\epsilon^{ijk}(\partial_{j}\partial^{l}v_{kl}-\nabla^{2}\partial_{j}G_{0k})+\frac{1}{\mu}\epsilon^{ijk}\partial_{j}\partial^{l}\Delta_{kl}-\nabla^{2}G_{0}{}^{i} + \partial^{i}\partial^{j}G_{0j} - \partial^{i}v^{j}{}_{j} + \partial_{j}v^{ij}
\nonumber \\
&& + \: [\eta^{kl}\partial^{i}-\frac{1}{2}(\eta^{ik}\partial^{l}+\eta^{il}\partial^{k})]\Delta_{kl} \approx 0,
\label{eq36}
\end{eqnarray}
where we observe that there are no more constraints because  (\ref{eq36}) gives relations for   the Lagrange multipliers. On the other hand, we can observe from the relations of the Lagrange multipliers  that $\partial_{j}\eqref{Multipliers1}=\eqref{eq36}$, thus, the process for identifying further constraints is finished. Therefore, the full set of constraints is given by 
\begin{eqnarray}
\nonumber
\varphi^{00}&\equiv&p^{00},
\\
\nonumber
\varphi^{0i}&\equiv&p^{0i},
\\
\nonumber
\varphi^{ij}&\equiv&p^{ij}+\frac{1}{2\mu}(\epsilon^{ikl}\eta^{jm}+\epsilon^{jkl}\eta^{im})\partial_{k}h_{lm},
\\
\nonumber
\Phi^{00}&\equiv&\pi^{00},
\\
\nonumber
\Phi^{0i}&\equiv&\pi^{0i},
\\
\nonumber
\Phi^{ij}&\equiv&\pi^{ij}+\frac{1}{\mu}(\epsilon^{ikl}\partial^{j}+\epsilon^{jkl}\partial^{i})\partial_{k}h_{0l}-\frac{1}{2\mu}(\epsilon^{ikl}\eta^{jm}+\epsilon^{jkl}\eta^{im})\partial_{k}G_{lm}
\\
\nonumber
& & - \: G^{ij}+(G^{k}{}_{k}-2\partial^{k}h_{0k})\eta^{ij}+(\partial^{i}h_{0}{}^{j}+\partial^{j}h_{0}{}^{i}),
\\
\nonumber
\Psi^{00}&\equiv&\nabla^{2}h^{i}{}_{i}-\partial_{i}\partial_{j}h^{ij},
\\
\nonumber
\Psi^{0i}&\equiv&\frac{1}{\mu}\epsilon^{ijk}(\partial_{j}\partial^{l}G_{kl}-\nabla^{2}\partial_{j}h_{0k})-\nabla^{2}h_{0}{}^{i}+\partial^{i}\partial^{j}h_{0j}-\partial^{i}G^{j}{}_{j}+\partial_{j}G^{ij}.
\label{Constraints}
\end{eqnarray}
With all constraints at hand, we need to classify them into first  and second class.  For this aim  we form the matrix $W^{IJ}$, whose entries are the Poisson brackets between all constraints, this is 
\begin{equation}
W^{IJ}=
\bordermatrix{
 & \varphi^{00} & \varphi^{0k} & \varphi^{kl} & \Phi^{00} & \Phi^{0k} & \Phi^{kl} & \Psi^{00} & \Psi^{0k} &  \cr
\varphi^{00} & 0 & 0 & 0 & 0 & 0 & 0 & 0 & 0 &  \cr
\varphi^{0i} & 0 & 0 & 0 & 0 & 0 & 0 & 0 & 0 &  \cr
\varphi^{ij} & 0 & 0 & 0 & 0 & 0 & \lbrace \varphi^{ij},\Phi^{kl} \rbrace & 0 & \lbrace \varphi^{ij},\Psi^{0k} \rbrace &  \cr
\Phi^{00} & 0 & 0 & 0 & 0 & 0 & 0 & 0 & 0 & \cr
\Phi^{0i} & 0 & 0 & 0 & 0 & 0 & \lbrace \Phi^{0i},\Phi^{kl} \rbrace & 0 & \lbrace \Phi^{0i},\Psi^{0k} \rbrace &  \cr
\Phi^{ij} & 0 & 0 & \lbrace \Phi^{ij},\varphi^{kl} \rbrace & 0 & \lbrace \Phi^{ij},\Phi^{0k} \rbrace & 0 & \lbrace \Phi^{ij},\Psi^{00} \rbrace & 0 &  \cr
\Psi^{00} & 0 & 0 & 0 & 0 & 0 & \lbrace \Psi^{00},\Phi^{kl} \rbrace & 0 & 0 &  \cr
\Psi^{0i} & 0 & 0 & \lbrace \Psi^{0i},\varphi^{kl} \rbrace & 0 & \lbrace \Psi^{0i},\Phi^{0k} \rbrace & 0 & 0 & 0 &  \cr}
\end{equation}
where $W^{IJ}$ is a 24$\times$24 matrix. The non-zero Poisson brackets  in $W^{IJ}$ are given by 
\begin{eqnarray}
\nonumber
\lbrace \varphi^{ij},\Phi^{kl} \rbrace &=& \left[\frac{1}{2\mu}(\epsilon^{ikm}\eta^{jl}+\epsilon^{jkm}\eta^{il}+\epsilon^{ilm}\eta^{jk}+\epsilon^{jlm}\eta^{ik})\partial_{m}+\frac{1}{2}(\eta^{ik}\eta^{jl}+\eta^{jk}\eta^{il})-\eta^{ij}\eta^{kl}\right]\delta^{3}(x-y),
\\
\nonumber
\lbrace \Phi^{0i},\Phi^{kl} \rbrace &=& \left[-\frac{1}{2\mu}(\epsilon^{ikm}\partial^{l}+\epsilon^{ilm}\partial^{k})\partial_{m}+\eta^{kl}\partial^{i}-\frac{1}{2}(\eta^{ik}\partial^{l}+\eta^{il}\partial^{k})\right]\delta^{3}(x-y),
\\
\nonumber
\lbrace \varphi^{ij},\Psi^{0k} \rbrace &=& \left[-\frac{1}{2\mu}(\epsilon^{ikl}\partial^{j}+\epsilon^{jkl}\partial^{i})\partial_{l} + \eta^{ij}\partial^{k}-\frac{1}{2}(\eta^{ik}\partial^{j}+\eta^{jk}\partial^{i})\right]\delta^{3}(x-y),
\\
\nonumber
\lbrace \Phi^{0i},\Psi^{0k} \rbrace &=& \left[\frac{1}{2\mu}\epsilon^{ikl}\partial_{l}\nabla^{2}+\frac{1}{2}\eta^{ik}\nabla^{2}-\frac{1}{2}\partial^{i}\partial^{k}\right]\delta^{3}(x-y),
\\
\lbrace \Phi^{ij},\Psi^{00} \rbrace &=& -\eta^{ij}\nabla^{2}\delta^{3}(x-y) + \partial^{i}\partial^{j}\delta^{3}(x-y),
\label{Parentesis(Parte3)}
\end{eqnarray}
Hence, after a long algebraic work, we find that the  matrix  $W^{IJ}$ has  12 null vectors, therefore,  this implies that there are   12 first class constraints \cite{19}, which  are  given by 
\begin{eqnarray}
\nonumber
\Gamma_{1}&\equiv&p^{00}\approx0,
\\
\nonumber
\Gamma_{2}&\equiv&\pi^{00}\approx0,
\\
\nonumber
\Gamma_{3}^{i}&\equiv&\pi^{0i}-\partial_{j}p^{ij}-\frac{1}{2\mu}\epsilon^{ijk}\partial_{j}\partial^{l}h_{kl}\approx0,
\\
\nonumber
\Gamma_{4}^{i}&\equiv&\partial_{j}\pi^{ij}+\frac{1}{2\mu}\epsilon^{ijk}\partial_{j}\partial^{l}G_{kl}\approx0,
\\
\nonumber
\Gamma_{5}^{i}&\equiv&p^{0i}\approx0,
\\
\label{FirstClassConstraints}
\Gamma_{6}&\equiv&\nabla^{2}h^{i}{}_{i}-\partial_{i}\partial_{j}h^{ij}+\partial_{i}\partial_{j}p^{ij}\approx0,
\end{eqnarray}
and the matrix has a rank=12, then the rank allows us to identify the following 12 second class constraints
\begin{eqnarray}
\nonumber
\chi_{1}^{ij}&\equiv&p^{ij}+\frac{1}{2\mu}(\epsilon^{ikl}\eta^{jm}+\epsilon^{jkl}\eta^{im})\partial_{k}h_{lm}\approx0,
\\
\nonumber
\chi_{2}^{ij}&\equiv&\pi^{ij}+\frac{1}{\mu}(\epsilon^{ikl}\partial^{j}+\epsilon^{jkl}\partial^{i})\partial_{k}h_{0l}-\frac{1}{2\mu}(\epsilon^{ikl}\eta^{jm}+\epsilon^{jkl}\eta^{im})\partial_{k}G_{lm}
\\
\label{SecondClassConstraints}
& & - \: G^{ij}+(G^{k}{}_{k}-2\partial^{k}h_{0k})\eta^{ij}+(\partial^{i}h_{0}{}^{j}+\partial^{j}h_{0}{}^{i})\approx0.
\end{eqnarray}
 The Poisson brackets between those  constraints are given by  
\begin{eqnarray}
\nonumber
\lbrace\Gamma^{a},\Gamma^{b}\rbrace &=& 0,
\\
\nonumber
\lbrace \chi_{1}^{ij},\chi_{1}^{kl} \rbrace &=& 0,
\\
\nonumber
\lbrace \chi_{1}^{ij},\chi_{2}^{kl} \rbrace &=& \left[\frac{1}{2\mu}(\epsilon^{ikm}\eta^{jl}+\epsilon^{jkm}\eta^{il}+\epsilon^{ilm}\eta^{jk}+\epsilon^{jlm}\eta^{ik})\partial_{m}+\frac{1}{2}(\eta^{ik}\eta^{jl}+\eta^{jk}\eta^{il})-\eta^{ij}\eta^{kl}\right]\delta^{3}(x-y),
\nonumber
\\
\label{Algebra}
\lbrace \chi_{2}^{ij},\chi_{2}^{kl} \rbrace &=& 0 .
\end{eqnarray}
In this manner, with all constraints classified into first and second class, we can perform the counting of physical degrees of freedom as follows: there are 40 canonical variables  $(h_{\mu\nu},G_{\mu\nu})$, the number of first class constraints is 12  and  there are 12 second class constraints, therefore the number of physical  degrees of freedom is $DOF=\frac{1}{2}[(40)-2(12)-12]=2$, just like the EH action. It is worth commenting  that those degrees of freedom are associated to two linearly independent polarizations of gravitational waves. \\
On the other hand, we can remove  the second class constraints by introducing the Dirac  brackets. In fact, the Dirac brackets are defined by 
\begin{equation}
\label{DiracBracket}
\lbrace F(x),G(y) \rbrace_{D} = \lbrace F(x),G(y) \rbrace - \int \lbrace F(x),\chi^{a}(u) \rbrace(C_{ab})^{-1}\lbrace \chi^{b}(v),G(y) \rbrace dudv,
\end{equation}
where $(C_{ab})^{-1}$ is the inverse of the matrix $C_{ab}$ whose entries are the Poisson brackets between the second  class constraints, this is 
\begin{equation}
\label{C}
C_{ab}=
\bordermatrix{
 & \chi_{1}^{kl} & \chi_{2}^{kl} \cr
\chi_{1}^{ij} & 0 & \lbrace \chi_{1}^{ij},\chi_{2}^{kl} \rbrace \cr
 \cr
\chi_{2}^{ij} & \lbrace \chi_{2}^{ij},\chi_{1}^{kl} \rbrace & 0 \cr
}
,
\end{equation}
the explicit form of $C_{ab}$ is given by 
\begin{equation}
\label{CExplicit}
C_{ab}=\frac{1}{\mu}
\bordermatrix{
 & \chi_{1}^{11} & \chi_{1}^{12} & \chi_{1}^{13} & \chi_{1}^{22} & \chi_{1}^{23} & \chi_{1}^{33} & \chi_{2}^{11} & \chi_{2}^{12} & \chi_{2}^{13} & \chi_{2}^{22} & \chi_{2}^{23} & \chi_{2}^{33} \cr
\chi_{1}^{11} & 0 & 0 & 0 & 0 & 0 & 0 & 0 & \partial_{3} & -\partial_{2} & -\mu & 0 & -\mu \cr
\chi_{1}^{12} & 0 & 0 & 0 & 0 & 0 & 0 & -\partial_{3} & \frac{\mu}{2} & \frac{1}{2}\partial_{1} & \partial_{3} & -\frac{1}{2}\partial_{2} & 0 \cr
\chi_{1}^{13} & 0 & 0 & 0 & 0 & 0 & 0 & \partial_{2} & -\frac{1}{2}\partial_{1} & \frac{\mu}{2} & 0 & \frac{1}{2}\partial_{3} & -\partial_{2} \cr
\chi_{1}^{22} & 0 & 0 & 0 & 0 & 0 & 0 & -\mu & -\partial_{3} & 0 & 0 & \partial_{1} & -\mu \cr
\chi_{1}^{23} & 0 & 0 & 0 & 0 & 0 & 0 & 0 & \frac{1}{2}\partial_{2} & -\frac{1}{2}\partial_{3} & -\partial_{1} & \frac{\mu}{2} & \partial_{1} \cr
\chi_{1}^{33} & 0 & 0 & 0 & 0 & 0 & 0 & -\mu & 0 & \partial_{2} & -\mu & -\partial_{1} & 0 \cr
\chi_{2}^{11} & 0 & -\partial_{3} & \partial_{2} & \mu & 0 & \mu & 0 & 0 & 0 & 0 & 0 & 0 \cr
\chi_{2}^{12} & \partial_{3} & -\frac{\mu}{2} & -\frac{1}{2}\partial_{1} & -\partial_{3} & \frac{1}{2}\partial_{2} & 0 & 0 & 0 & 0 & 0 & 0 & 0 \cr
\chi_{2}^{13} & -\partial_{2} & \frac{1}{2}\partial_{1} & -\frac{\mu}{2} & 0 & -\frac{1}{2}\partial_{3} & \partial_{2} & 0 & 0 & 0 & 0 & 0 & 0 \cr
\chi_{2}^{22} & \mu & \partial_{3} & 0 & 0 & -\partial_{1} & \mu & 0 & 0 & 0 & 0 & 0 & 0 \cr
\chi_{2}^{23} & 0 & -\frac{1}{2}\partial_{2} & \frac{1}{2}\partial_{3} & \partial_{1} & -\frac{\mu}{2} & -\partial_{1} & 0 & 0 & 0 & 0 & 0 & 0 \cr
\chi_{2}^{33} & \mu & 0 & -\partial_{2} & \mu & \partial_{1} & 0 & 0 & 0 & 0 & 0 & 0 & 0 \cr
}
\delta^{3}(x-y)
,
\end{equation}
we can observe that this matrix is not singular, therefore it has an  inverse. Hence,  after long algebraic manipulations, we find  the following non-trivial Dirac brackets 
\begin{eqnarray}
\nonumber
\lbrace h_{ij},\pi^{kl} \rbrace_{D} &=& \frac{1}{2}(\delta_{i}^{k}\delta_{j}^{l}+\delta_{i}^{l}\delta_{j}^{k})\delta^{3}(x-y) + \frac{\mu^{2}}{4\Xi}\left[[(\delta_{i}^{k}\delta_{j}^{l}+\delta_{i}^{l}\delta_{j}^{k}- \eta_{ij}\eta^{kl})\nabla^{2} + (\eta_{ij}\partial^{k}\partial^{l}+\eta^{kl}\partial_{i}\partial_{j})](\nabla^{2}+\mu^{2})\right.
\\
\nonumber
&& - 3\partial_{i}\partial_{j}\partial^{k}\partial^{l} - \frac{3\mu^{2}}{4}(\delta_{i}^{k}\partial_{j}\partial^{l}+\delta_{i}^{l}\partial_{j}\partial^{k}+\delta_{j}^{k}\partial_{i}\partial^{l}+\delta_{j}^{l}\partial_{i}\partial^{k}) +\frac{\mu}{4}\left[(\epsilon_{i}{}^{km}\delta_{j}^{l}+\epsilon_{j}{}^{km}\delta_{i}^{l}+\epsilon_{i}{}^{lm}\delta_{j}^{k}\right.
\\
\label{DB1}
&& \left. +\epsilon_{j}{}^{lm}\delta_{i}^{k})(\nabla^{2}+\mu^{2}) + 3(\epsilon_{i}{}^{km}\partial_{j}\partial^{l}+\epsilon_{j}{}^{km}\partial_{i}\partial^{l}+\epsilon_{i}{}^{lm}\partial_{j}\partial^{k}+\epsilon_{j}{}^{lm}\partial_{i}\partial^{k})]\partial_{m}\right]\delta^{3}(x-y),
\\
\nonumber
\\
\nonumber
\lbrace G_{ij},p^{kl} \rbrace_{D} &=& - \frac{\mu^{2}}{4\Xi}\left[[(\delta_{i}^{k}\delta_{j}^{l}+\delta_{i}^{l}\delta_{j}^{k}- \eta_{ij}\eta^{kl})\nabla^{2} + (\eta_{ij}\partial^{k}\partial^{l}+\eta^{kl}\partial_{i}\partial_{j})](\nabla^{2}+\mu^{2}) - 3\partial_{i}\partial_{j}\partial^{k}\partial^{l}\right.
\\
\nonumber
&& - \frac{3\mu^{2}}{4}(\delta_{i}^{k}\partial_{j}\partial^{l}+\delta_{i}^{l}\partial_{j}\partial^{k}+\delta_{j}^{k}\partial_{i}\partial^{l}+\delta_{j}^{l}\partial_{i}\partial^{k}) + \frac{\mu}{4}[(\epsilon_{i}{}^{km}\delta_{j}^{l}+\epsilon_{j}{}^{km}\delta_{i}^{l}+\epsilon_{i}{}^{lm}\delta_{j}^{k}+\epsilon_{j}{}^{lm}\delta_{i}^{k})(\nabla^{2}+\mu^{2})
\\
\label{DB2}
&& \left. + 3(\epsilon_{i}{}^{km}\partial_{j}\partial^{l}+\epsilon_{j}{}^{km}\partial_{i}\partial^{l}+\epsilon_{i}{}^{lm}\partial_{j}\partial^{k}+\epsilon_{j}{}^{lm}\partial_{i}\partial^{k})]\partial_{m}\right]\delta^{3}(x-y),
\end{eqnarray}

\begin{eqnarray}
\nonumber
\lbrace \pi^{ij},p^{kl} \rbrace_{D} &=& \frac{1}{8\mu}(\epsilon^{ikm}\eta^{jl}+\epsilon^{jkm}\eta^{il}+\epsilon^{ilm}\eta^{jk}+\epsilon^{jlm}\eta^{ik})\partial_{m}\delta^{3}(x-y) - \frac{\mu^{2}}{8\Xi}\left[[(\eta^{ik}\eta^{jl}+\eta^{il}\eta^{jk}- \eta^{ij}\eta^{kl})\nabla^{2}\right.
\\
\nonumber
&& \left. + (\eta^{ij}\partial^{k}\partial^{l}+\eta^{kl}\partial^{i}\partial^{j})\right](\nabla^{2}+\mu^{2}) - 3\partial^{i}\partial^{j}\partial^{k}\partial^{l} - \frac{3\mu^{2}}{4}(\eta^{ik}\partial^{j}\partial^{l}+\eta^{il}\partial^{j}\partial^{k}+\eta^{jk}\partial^{i}\partial^{l}+\eta^{jl}\partial^{i}\partial^{k})
\\
\nonumber
&& + \frac{\mu}{4}[(\epsilon^{ikm}\eta^{jl}+\epsilon^{jkm}\eta^{il}+\epsilon^{ilm}\eta^{jk}+\epsilon^{jlm}\eta^{ik})(\nabla^{2}+\mu^{2}) + 3(\epsilon^{ikm}\partial^{j}\partial^{l}+\epsilon^{jkm}\partial^{i}\partial^{l}
\\
\label{DB3}
&& \left. +\epsilon^{ilm}\partial^{j}\partial^{k}+\epsilon^{jlm}\partial^{i}\partial^{k})]\partial_{m}\right]\delta^{3}(x-y),
\\
\nonumber
\\
\nonumber
\lbrace h_{ij},G_{kl} \rbrace_{D} &=& \frac{1}{2}(\eta_{ik}\eta_{jl}+\eta_{il}\eta_{jk}-\eta_{ij}\eta_{kl})\delta^{3}(x-y) + \frac{\mu^{2}}{2\Xi}\left[[(\eta_{ik}\eta_{jl}+\eta_{il}\eta_{jk}- \eta_{ij}\eta_{kl})\nabla^{2} + (\eta_{ij}\partial_{k}\partial_{l}\right.
\\
\nonumber
&& \left. +\eta_{kl}\partial_{i}\partial_{j})\right](\nabla^{2}+\mu^{2}) - 3\partial_{i}\partial_{j}\partial_{k}\partial_{l} - \frac{3\mu^{2}}{4}(\eta_{ik}\partial_{j}\partial_{l}+\eta_{il}\partial_{j}\partial_{k}+\eta_{jk}\partial_{i}\partial_{l}+\eta_{jl}\partial_{i}\partial_{k})
\\
\nonumber
&& + \frac{\mu}{4}[(\epsilon_{ik}{}^{m}\eta_{jl}+\epsilon_{jk}{}^{m}\eta_{il}+\epsilon_{il}{}^{m}\eta_{jk}+\epsilon_{jl}{}^{m}\eta_{ik})(\nabla^{2}+\mu^{2}) + 3(\epsilon_{ik}{}^{m}\partial_{j}\partial_{l}+\epsilon_{jk}{}^{m}\partial_{i}\partial_{l}
\\
\label{DB4}
&& \left. +\epsilon_{il}{}^{m}\partial_{j}\partial_{k}+\epsilon_{jl}{}^{m}\partial_{i}\partial_{k})]\partial_{m}\right]\delta^{3}(x-y),
\\
\nonumber
\\
\label{DB5}
\lbrace G_{ij},\pi^{0k} \rbrace_{D} &=& - \frac{1}{2}(\delta_{i}^{k}\partial_{j}+\delta_{j}^{k}\partial_{i})\delta^{3}(x-y),
\\
\nonumber
\\
\label{DB6}
\lbrace \pi^{ij},\pi^{0k} \rbrace_{D} &=& \frac{1}{4\mu}(\epsilon^{ikl}\partial^{j}+\epsilon^{jkl}\partial^{i})\partial_{l}\delta^{3}(x-y),
\end{eqnarray}
where $\Xi=-\mu^{2}(\nabla^{2}+\mu^{2})(\nabla^{4}+\frac{\mu^{2}}{4})$. We can see a  contribution due to the CS term which   could be important in the analysis of quantization of the theory.  Furthermore,  with this new set of brackets one can calculate  the Dirac algebra between the first class constraints and the second class constraints, the algebra is  given by 
\begin{eqnarray}
\lbrace \Gamma^{a},\Gamma^{b} \rbrace_{D} &=& 0,
\\
\lbrace \Gamma^{a},\chi^{b} \rbrace_{D} &=& 0.
\end{eqnarray}
Once we have  introduced the Dirac brackets  we proceed to identify the symmetries of the theory, for instance,  the gauge transformations. For this part  we will use the first class constraints and the Dirac brackets calculated above. Hence, the local gauge transformations are given  by 
\begin{equation}
\label{Variation}
\delta X(x)=\int \lbrace X(x),\omega_{a}\Gamma^{a}(y)\rbrace_{D}d^{3}y ,
\end{equation}
where $\omega_{\alpha}$ are gauge parameters. From the calculation of  \eqref{Variation} for the perturbation $h_{\alpha \beta}$ and  the $G_{\alpha \beta} $ variables,   we obtain
\begin{eqnarray}
\delta h_{00} &=& \omega_{2}, \nonumber 
\\
\delta h_{0i} &=& \frac{1}{2}\omega_{3i},\nonumber 
\\
\delta h_{ij} &=& -\frac{1}{2}\left( \partial_{i}\omega_{4j}+\partial_{j}\omega_{4i} \right),  \nonumber 
\\
\delta G_{00} &=& \omega_{1}, \nonumber 
\\
\delta G_{0i} &=& \frac{1}{2}\omega_{5i},  \nonumber 
\\
\delta G_{ij} &=& \frac{1}{2}(\partial_{i}\omega_{3j}+\partial_{j}\omega_{3i}) + \partial_{i}\partial_{j}\omega_{6}.
\label{eq55a}
\end{eqnarray}
It is worth commenting, that the gauge paremeters $\omega's$ are restricted. In fact, the involution relations of all  first class constraints provide equations for unfree gauge transformations, say, the gauge parameters obey a system of  partial differential equations \cite{Abakumova}. Thus, the involution relations between the constraints and Hamiltonian reads 
\begin{eqnarray}
\lbrace \Gamma_{1},{H'}_{can} \rbrace_{D} &=& - \pi^{00} = - \Gamma_{2},
\nonumber \\
\lbrace \Gamma_{5}^{i},{H'}_{can} \rbrace_{D} &=& - \pi^{0i} = \partial_{j}\chi_{1}^{ij} - \Gamma_{3}^{i}, \nonumber 
\\
\lbrace \Gamma_{2},{H'}_{can} \rbrace_{D} &=& \nabla^{2}h^{i}{}_{i} - \partial_{i}\partial_{j}h^{ij} = \Gamma_{6} - \partial_{i}\partial_{j}\chi_{1}^{ij}, \nonumber 
\\
\lbrace \Gamma_{3}^{i},{H'}_{can} \rbrace_{D} &=& \partial_{j}\pi^{ij} + \frac{1}{2\mu}\epsilon^{ijk}\partial_{j}\partial^{l}G_{kl} = \Gamma_{4}^{i}, \nonumber 
\\
\lbrace \Gamma_{4}^{i},{H'}_{can} \rbrace_{D} &=& 0,\nonumber 
\\
\lbrace \Gamma_{6},{H'}_{can} \rbrace_{D} &=& - \partial_{i}\partial_{j}\pi^{ij} = - \partial_{i}\Gamma_{4}^{i}, \nonumber 
\\
\lbrace \chi_{1}^{ij},{H'}_{can} \rbrace_{D} &=& 0, \nonumber 
\\
\lbrace \chi_{2}^{ij},{H'}_{can} \rbrace_{D} &=& 0, 
\label{eq61a}
\end{eqnarray}
where ${H'}_{can}$ is given in (\ref{CanonicalHamiltonian2}). Now, the identification of  structure coefficients given in (\ref{eq61a}) and the general Noether relations reported in \cite{Abakumova}  allow us identify the constraints on gauge parameters $\omega_{a}$,   given by 
\begin{eqnarray}
\dot{\varepsilon}^{a+1} + O^{a}_{b}\varepsilon^{b} =0, 
\end{eqnarray}
then we have
\begin{eqnarray}
\nonumber
&& \dot{\varepsilon}^{\Gamma_{2}} + O_{\Gamma_{1}}^{\Gamma_{2}}\omega^{1} =0, \nonumber 
\\
&\Rightarrow& \int[\partial_{0}\omega_{2}\delta^{3}(x-y) - \delta^{3}(x-y)\omega_{1} ]d^{3}y =0, \nonumber 
\\
\nonumber
&& \dot{\varepsilon}^{\Gamma_{3}^{i}} + O_{\Gamma_{5}^{i}}^{\Gamma_{3}^{i}}\varepsilon^{\Gamma_{5}^{i}} =0, \nonumber 
\\
&\Rightarrow& \int[ \partial_{0}\omega_{3i}\delta^{3}(x-y) - \omega_{5i}\delta^{3}(x-y) ]d^{3}y =0, \nonumber 
\\
\nonumber
&& \dot{\varepsilon}^{\Gamma^{6}} + O_{\Gamma_{2}}^{\Gamma_{6}}\varepsilon^{\Gamma_{2}} =0, \nonumber 
\\
&\Rightarrow& \int[ \partial_{0}\omega_{6}\delta^{3}(x-y) + \omega_{2}\delta^{3}(x-y) ]d^{3}y =0, \nonumber 
\\
\nonumber
&& \dot{\varepsilon}^{\Gamma_{4}^{i}} + O_{\Gamma_{3}^{i}}^{\Gamma_{4}^{i}}\varepsilon^{\Gamma_{3}^{i}} + O^{\Gamma_{4}^{i}}_{\Gamma_{6}}\varepsilon^{\Gamma_{6}} =0, \nonumber  
\\
&\Rightarrow& \int[ \partial_{0}\omega_{i}^{4}\delta^{3}(x-y) + \omega_{3i}\delta^{3}(x-y) - \partial_{i}\delta^{3}(x-y)\omega_{6}]d^{3}y =0,
\label{eq63}
\end{eqnarray}
thus, from (\ref{eq63}) we obtain an explicit set of relations between the gauge parameters in the following form 
\begin{eqnarray}
&& \partial_{0}\omega_{2} - \omega_{1} = 0, \nonumber 
\\
&& \partial_{0}\omega_{3i} - \omega_{5i} = 0, \nonumber 
\\
&& \partial_{0}\omega_{6} + \omega_{2} = 0, \nonumber 
\\
&& \partial_{0}\omega_{4i} + \omega_{3i} + \partial_{i}\omega_{6} = 0. 
\label{eq59a}
\end{eqnarray}
Afterwards, by using the relations (\ref{eq59a}) into (\ref{eq55a})  we can rewrite the gauge transformations for $h_{\mu\nu}$ and $G_{\mu\nu}$ in the following way
\begin{eqnarray}
\label{dh00}
\delta h_{00} &=& - \partial_{0}\omega_{6},
\\
\label{dh0i}
\delta h_{0i} &=& - \frac{1}{2}(\partial_{0}\omega_{4i}+\partial_{i}\omega_{6}),
\\
\label{dhij}
\delta h_{ij} &=& - \frac{1}{2}(\partial_{i}\omega_{4j}+\partial_{j}\omega_{4i}),
\\
\label{dG00}
\delta G_{00} &=& -\partial_{0}\partial_{0}\omega_{6},
\\
\label{dG0i}
\delta G_{0i} &=& - \frac{1}{2}\partial_{0}(\partial_{0}\omega_{4i}+\partial_{i}\omega_{6}),
\\
\label{dGij}
\delta G_{ij} &=& - \frac{1}{2}(\partial_{i}\partial_{0}\omega_{4j}+\partial_{j}\partial_{0}\omega_{4i}),
\end{eqnarray}
moreover, we can express \eqref{dh00}-\eqref{dGij} in a covariant way by introducing  the following   $\omega_{6}=-2\Lambda_{0}$ and $\omega_{4i}=-2\Lambda_{i}$, thus, the gauge transformations reads
\begin{eqnarray}
\label{trans}
\delta h_{\mu\nu} &=& \partial_{\mu}\Lambda_{\nu} + \partial_{\nu}\Lambda_{\mu},
\\
\delta G_{\mu\nu} &=& \partial_{0}(\partial_{\mu}\Lambda_{\nu} + \partial_{\nu}\Lambda_{\mu}).
\end{eqnarray}
The action (\ref{I}) is invariant under the transformations (\ref{trans}). In fact, we observe that (\ref{eqlin}) transforms  like $\delta{G}_{\mu\nu}^{lin}=0$  and $\delta{C}_{\mu\nu}^{lin}=0$. Furthermore, we observe that $\partial^\mu G_{\mu\nu}^{lin}=0$ and  $\partial^\mu C_{\mu\nu}^{lin}=0$, therefore   
\begin{equation}
\delta S[h_{\mu\nu}]=0.
\end{equation}

\section{Conclusions}
In this paper, a detailed GLT analysis for JY theory has been performed. Using the null vectors, we have reported the complete structure of the constraints. Furthermore, with the correct constraints at hand, the counting of physical degrees of freedom was performed. We identified two physical degrees of freedom, thus, the EH and JY theories share either the same classical test of gravity or the number of physical degrees of freedom. Despite this similarity, both theories present significant differences at the fundamental  Dirac brackets level due to the modification, and of course, if we take the standard limit on $\mu \rightarrow \infty$, then the usual Dirac brackets of EH theory are recovered \cite{17, Escalante}. Moreover, the Dirac brackets were helpfull in seeing whether Ostrogradski's  instabilities are present or not, this can be seen in the appendix B. In fact, with the Dirac brackets at hand, we used the second class constraints for eliminating the linear momenta terms in the Hamiltonian. In addition, Dirac's brackets  were also used to find the involution relations between the constraints, as well as the unfree gauge symmetry. It is worth mentioning that the Dirac brackets are relevant  in the analysis of quantum aspects, like  the calculation of the propagators of the fields. However,  the calculations of the propagators is a dificult task when   higher derivatives are present  \cite{24}. In any case, these ideas are being analyzed in the non-perturbative and perturbative sector,  they are still in progress and we will utilize this work as a base for the subject of forthcoming works \cite{23}.

\subsection*{Appendix A: Gitman-Lyakhovich-Tyutin approach}

In general, a higher-order Lagrangian has the form
\begin{equation}
\label{LAppx}
\mathcal{L} = \mathcal{L}(h_{\mu\nu},\partial_{i}h_{\mu\nu},\partial_{i}\partial_{j}h_{\mu\nu},\dot{h}_{\mu\nu},\partial_{i}\dot{h}_{\mu\nu},\ddot{h}_{\mu\nu})
\end{equation}
the equations of motion that arise from \eqref{LAppx} are given by
\begin{eqnarray}
\label{EoMAppx}
\frac{\partial \mathcal{L}}{\partial h_{\mu\nu}} - \partial_{0}\frac{\partial \mathcal{L}}{\partial\dot{h}_{\mu\nu}} - \partial_{i}\frac{\partial \mathcal{L}}{\partial(\partial_{i}h_{\mu\nu})} + \partial_{0}\partial_{0} \frac{\partial \mathcal{L}}{\partial(\ddot{h}_{\mu\nu})} + \partial_{0}\partial_{i}\frac{\partial \mathcal{L}}{\partial(\partial_{i}\dot{h}_{\mu\nu})} + \partial_{i}\partial_{j}\frac{\partial \mathcal{L}}{\partial(\partial_{i}\partial_{j}h_{\mu\nu})} = 0,
\end{eqnarray}
by defining the canonical momenta one can rewrite the action as
\begin{equation}
\mathcal{S}' = \int[ \mathcal{L} + \pi^{\mu}(\dot{h}_{\mu\nu}-G_{\mu\nu}) + p^{\mu\nu}(\dot{G}_{\mu\nu}-v_{\mu\nu}) ]d^{4}x = \int\mathcal{L}'d^{4}x
\end{equation}
where the dependence of the lagrangian $\mathcal{L}$ is $\mathcal{L}=\mathcal{L}(h_{\mu\nu},G_{\mu\nu},\partial_{i}h_{\mu\nu},v_{\mu\nu},\partial_{i}G_{\mu\nu},\partial_{i}\partial_{j}h_{\mu\nu})$ and the fields $G_{\mu\nu}=\dot{h}_{\mu\nu}$ and $v_{\mu\nu}=\dot{G}_{\mu\nu}$ are treated as independent canonical variables. Hence,  their respective equations of motions are:
\begin{eqnarray}
\label{EoM1Appx}
\frac{\delta \mathcal{S}'}{\delta h_{\mu\nu}} &=& \frac{\partial \mathcal{L}}{\partial h_{\mu\nu}} - \partial_{i}\frac{\partial \mathcal{L}}{\partial(\partial_{i}h_{\mu\nu})} + \partial_{i}\partial_{j}\frac{\partial \mathcal{L}}{\partial(\partial_{i}\partial_{j}h_{\mu\nu})} - \dot{\pi}^{\mu\nu} = 0,
\\
\label{EoM2Appx}
\frac{\delta \mathcal{S}'}{\delta G_{\mu\nu}} &=& \frac{\partial \mathcal{L}}{\partial G_{\mu\nu}} - \partial_{i}\frac{\partial \mathcal{L}}{\partial(\partial_{i}G_{\mu\nu})} - \pi^{\mu\nu} - \dot{p}^{\mu\nu} = 0,
\\
\label{EoM3Appx}
\frac{\delta \mathcal{S}'}{\delta v_{\mu\nu}} &=& \frac{\partial \mathcal{L}}{\partial v_{\mu\nu}} - p^{\mu\nu} = 0,
\\
\label{EoM4Appx}
\frac{\delta \mathcal{S}'}{\delta \pi^{\mu\nu}} &=& \dot{h}_{\mu\nu} - G_{\mu\nu} = 0,
\\
\label{EoM5Appx}
\frac{\delta \mathcal{S}'}{\delta p^{\mu\nu}} &=& \dot{G}_{\mu\nu} - v_{\mu\nu} = 0.
\end{eqnarray}

The equations \eqref{EoM1Appx}-\eqref{EoM5Appx} are equivalent to \eqref{EoMAppx}, moreover, in a regular system the equations \eqref{EoM3Appx} give explicit expressions for $v_{\mu\nu}$ in terms of the other canonical variables. For  singular theories  not all $v_{\mu\nu}$ can be determined and relations between the phase space variables are established i.e. the primary constraints of the theory emerge. At this point one can follow all  Dirac steps,  by demanding consistency on the primary constraints and so on.

\subsection*{Appendix B: Ostrogradski's instability}

It is known in the literature that if a Lagrangian involves higher-order time-derivative terms, then the Hamiltonian of the system could depend linearly on a canonical momenta \cite{Ostrogradski}, this means that it has no local minimum and its energy will unbounded from below. In our analysis the hamiltonian under study  depends linearly on $\pi^{\mu\nu}$ and $p^{\mu\nu}$ (see \ref{CanonicalHamiltonian}) and it could suffer from the sickness of Ostrogradski's instability. Nonetheless, it is well-known that the instability can be healed if there are constraints \cite{Ganz}. In fact, we have constructed the Dirac brackets, then the second class constraints can be considered strongly zero. Hence, the second class constraints can be used to rewrite \eqref{CanonicalHamiltonian}, thus, after a long algebraic work, the Hamiltonian (\ref{CanonicalHamiltonian}) takes the form
\begin{eqnarray}
\nonumber
H_{can}' &=& \frac{1}{2}\pi^{ij}\pi_{ij} - \frac{1}{4}\pi^{i}{}_{i}\pi^{j}{}_{j} + 2\partial^{i}h_{0i}\partial^{j}h_{0j} - 2\partial^{i}h_{0}{}^{j}\partial_{i}h_{0j} + 2\partial_{i}h_{0j}G^{ij} - 2\partial^{i}h_{0i}G^{j}{}_{j} + \frac{1}{2}\partial^{k}h^{ij}\partial_{k}h_{ij}
\\
\nonumber
&& - \frac{1}{2}\partial_{k}h^{i}{}_{i}\partial^{k}h^{j}{}_{j} + \partial_{i}h^{ij}\partial_{j}h^{k}{}_{k} - \partial_{k}h^{k}{}_{i}\partial_{j}h^{ij} - \frac{1}{\mu}\epsilon^{ijk}[\partial^{l}h_{im}\partial^{m}\partial_{j}h_{kl} - 3 \nabla^{2}h_{0i}\partial_{j}h_{0k}
\\
\nonumber
&& + \nabla^{2}h_{i}{}^{l}\partial_{j}h_{kl} + 3\partial_{i}h_{0j}\partial^{l}G_{kl}]  + \frac{1}{\mu^{2}}[(\partial^{i}\partial^{j}h_{0j}-\nabla^{2}h_{0}{}^{i})\nabla^{2}h_{0i} -(\partial^{i}\partial^{j}h_{0j}-\nabla^{2}h_{0}{}^{i})\partial^{k}G_{ik}]
\\
\label{CanonicalHamiltonian2}
&& + \frac{1}{4\mu^{2}}[ 2\partial^{i}G_{ij}\partial^{j}G^{k}{}_{k} - \partial^{k}G^{i}{}_{k}\partial^{j}G_{ij} - \partial_{k}G^{i}{}_{i}\partial^{k}G^{j}{}_{j}]
\end{eqnarray}
where we observe that there are no linear terms on the momenta anymore. Now, with the Hamiltonian (\ref{CanonicalHamiltonian2}) we can calculate the equations of motions, for example, that for $\pi_{0i}$, this is
\begin{equation}
\nonumber
\dot{\pi}_{0i} = \lbrace \pi_{0i},H_{can}' \rbrace_{D} = \nabla^{2}h_{0i} - \partial_{i}\partial^{j}h_{0j} + \partial_{i}G^{j}{}_{j} - \partial^{j}G_{ij} - \frac{1}{\mu}\epsilon_{i}{}^{jk}(\partial_{j}\partial^{l}G_{kl} - \nabla^{2}\partial_{j}h_{0k}) = \lbrace \pi_{0i},H_{can} \rbrace.
\end{equation}
where we observe that these equations of motion are in agreement with the evolution of  the Dirac and Poisson brackets \cite{Ganz}


\end{document}